\documentclass[twocolumn,showpacs,preprintnumbers,amsmath,amssymb]{revtex4}
\usepackage{amssymb}     
\usepackage{graphicx}    
\usepackage{dcolumn}
\usepackage{bm}
\usepackage{color}

\begin{document}

\title{Stochastic Gravitational Wave Background originating from Halo Mergers}

\author{Takahiro Inagaki$^1$, Keitaro Takahashi$^{2,1}$, and
Naoshi Sugiyama$^{1,3,4}$}
       
\affiliation{
$^1$Department of Physics and Astrophysics, Nagoya University,
Nagoya 464-8602, Japan; inagaki.takahiro@b.nagoya-u.jp\\
$^2$Faculty of Science, Kumamoto University, 2-39-1, Kurokami, Kumamoto 860-8555, Japan\\
$^3$Institute for the Physics and Mathematics of the Universe (IPMU),
The University of Tokyo, Kashiwa, Chiba, 277-8568, Japan\\
$^4$Kobayashi-Maskawa Institute for the Origin of Particles and the Universe,
Nagoya University, Nagoya 464-8602, Japan}

\date{\today}

\begin{abstract}

The stochastic gravitational wave background (GWB) from halo mergers
is investigated by a quasi-analytic method.  
The method we employ consists of two steps.  The first step is to 
construct a merger tree by using the Extended Press-Schechter
formalism or the Sheth \& Tormen formalism, 
with Monte-Carlo realizations.  
This merger tree provides evolution of halo masses. 
From $N$-body simulation of two-halo mergers, we can estimate 
the amount of gravitational wave emission induced by the individual merger
process.  
Therefore the second step is to combine this gravitaional wave emission to 
the merger tree and obtain the amplitude of GWB.  
We find $\Omega_{GW}\sim 10^{-19}$ for $f\sim
10^{-17}-10^{-16}$ Hz, where $\Omega_{GW}$ is the energy density of
the GWB.  It turns out that most of the contribution on the GWB comes from 
halos with masses 
below $10^{15} M_\odot$ and mergers at low redshift, i.e., $0<z<0.8$.  

\end{abstract}


\maketitle

\section{Introduction}
\label{introduction}

Stochastic gravitational wave background (GWB) is a valuable source
of information for both cosmology and astrophysics.

In the very early universe, inflation \cite{1981MNRAS.195..467S,
PhysRevD.23.347, PhysRevLett.48.1220, 1982PhLB..108..389L} is expected
to be the most feasible scenario.  It is believed that both scalar and
tensor mode perturbations from quantum fluctuations are produced in
the epoch of inflation.  The former evolve into large scale structure
of the universe.  And the latter directly travel toward us and are
observed as the GWB.  Therefore this GWB can be a direct probe of
inflation.  For that, one needs to solve the transfer of the
gravitational wave in the history of the expanding universe from the
epoch of inflation to present.  For example, 
\citet{2009PhRvD..79j3501K} studied evolution of the gravitational waves
from inflation in detail and showed
that the amplitude of the GWB energy spectrum 
is $\sim 10^{-15}$ for $f>10^{-17} {\rm Hz}$ while the
value is very dependent on models of inflation.

The GWB is also produced by second-order scalar perturbations.  In the
linear perturbations, it is known that structure formation does not
generate any gravitaional waves.  It turns out that, however,
gravitational waves with cosmological scales would be emitted during
the process of structure formation if one includes the second-order
terms in the perturbation equations as is studied in
\cite{2004PhRvD..69f3002M,2007PhRvD..75l3518A,2007PhRvD..76h4019B}.
The result obtained by them is that the amplitude of the GWB is
$10^{-20}-10^{-15}$ for a wide frequency range,

The best method for detecting such GWB with cosmological scales is 
to measure the B-mode of the Cosmic Microwave Background (CMB). 
More specifically, the CMB photons are very sensitive to 
gravitational waves with $f\sim 10^{-17} {\rm Hz}$ \cite{2005AnPhy.318....2P}. 
In actuality, several missions suited
for this aim have been planned; for example,
ACTPol, SPTpol, POLARBeaR, LiteBIRD (see, http://cmbpol.kek.jp/index-e.html),
and Cosmic Inflation Probe (see, http://www.cfa.harvard.edu/cip/).
Their primary purpose is to detect primordial gravitational waves
generated during inflation. 

In this paper, we investigate the GWB originating from dark halo
mergers, which are expected to produce gravitational waves with
cosmological scales.  In a hierarchical model of structure formation,
it is expected that low-mass dark halos repeatedly merge with each
other, then more massive dark halos are formed (see, e.g.,
\cite{1993MNRAS.262..627L,1994MNRAS.271..676L}).  A large amounts of
gravitational waves from the process would be emitted because the
process is a highly nonlinear event.  In \cite{2010PhRvD..82l4007I},
we studied the gravitational waves emitted from a single galaxy merger
with N-body simulations. The peak luminosity and total emitted energy were
found to reach about $10^{31}~{\rm erg/sec}$ and $10^{47}~{\rm erg}$,
respectively for a collision of two galaxies with masses
$3.8 \times 10^{12} M_{\odot}$. We also studied the relative
contribution of the disk, bulge and halo, the effect of initial
velosity and relative angular momentum.  To calculate the GWB,
we sum the gravitational wave spectrum from a single merger over
the merger history.

This paper is organized as follows: 
in section 2 our method for calculating the GWB is described; 
in section 3 we show the results; 
and in section 4 we provide summary and discussion.
In this paper we consider a spatially-flat CDM model with the following
cosmological parameters:
$\Omega_m = 0.275, \Omega_\Lambda = 0.725, h = 0.702,
\sigma_8 = 0.815$ and $n_s = 0.963$ for the density parameters
of matter and cosmological constant, the hubble constant, the density
fluctuation in sphere of $8 h^{-1}~{\rm Mpc}$ and the spectral
index of the primordial scalar fluctuations \cite{2011ApJS..192...18K}.

\section{Method}
\label{method}

In this section we describe our quasi-analytic method.
As the first step, we construct merger history
using Monte-Carlo simulation based on the method of \cite{1999MNRAS.305....1S}.
For the mass function of dark halos, we consider Extended Press-Schechter
formalism (EPS) \cite{1993MNRAS.262..627L,1994MNRAS.271..676L} and
Sheth \& Tormen formalism (ST) \cite{1999MNRAS.308..119S,2001MNRAS.321..372J},
although the former is mostly used below. In the EPS formalism which provides
the conditional mass function, the fraction of the trajectories in haloes
with mass $M_1$ at $z_1$ that are in halos with mass
$M_0$ at $z_0$ $(M_1<M_0,~z_0<z_1)$, is given by,
\begin{eqnarray}
&&f_{\rm PS}(M_1,z_1| M_0,z_0)d\sigma_1^2=\nonumber\\ 
&&\frac{1}{\sqrt{2\pi}}\frac{\delta_{c,1}-\delta_{c,0}}
{( \sigma_1^2-\sigma_0^2)^{3/2}}\exp\left[ 
-\frac{( \delta_{c,1}-\delta_{c,0} )^2}{2(\sigma_1^2-\sigma_0^2  )}\right]
d\sigma_1^2 ,
\label{eq:cf}
\end{eqnarray}
where $\delta_{c,i}$ and $\sigma_j$ are the threshold in the Press-Schechter
formalism at time $z_i$ and the mass variance for mass $M_j$, respectively. 

On the other hand, in case of ST, conditional mass function is described by,
\begin{eqnarray}
&&f_{\rm ST}(M_1,z_1| M_0,z_0)d\sigma_1^2=\nonumber\\ 
&&A\sqrt{\frac{a}{2\pi}}\frac{\delta_{c,1}-\delta_{c,0}}
{( \sigma_1^2-\sigma_0^2)^{3/2}}
\left[1+\left(\frac{\sigma^2_1-\sigma^2_0}{a(\delta_{c,1}-\delta_{c,0})^2}
\right)^p\right]\nonumber\\
&&\times\exp\left[-\frac{a( \delta_{c,1}-\delta_{c,0} )^2}{2(\sigma_1^2-\sigma_0^2  )}\right]
d\sigma_1^2 ,
\label{eq:cf2}
\end{eqnarray}
with $A=0.3222$, $a=0.707$ and $p=0.3$.

\begin{figure}[t]
\includegraphics[width=77mm]{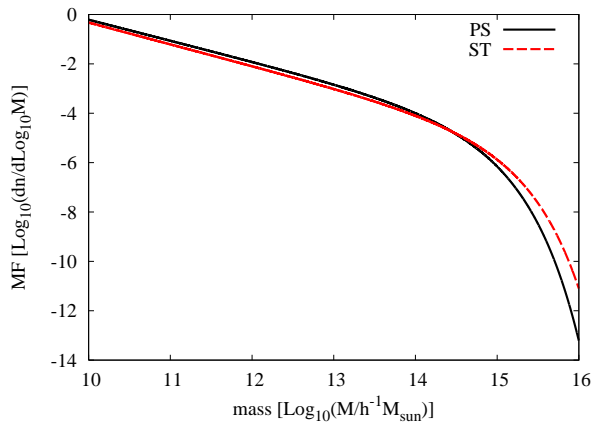}
\caption{The mass function at present by using EPS(solid) and ST(dashed). 
The y-axis is the number density in the unit of $1/({\rm h^{-1}Mpc})^3$.
The x-axis is the halo mass, $M$.}
\label{fig:MF}
\end{figure}

By using Eqs. (\ref{eq:cf}) and (\ref{eq:cf2}), the probability that is
a parent halo with mass $M_1$ at $z_1$ being in a progenitor halo
with mass $M_0$ at $z_0$ is obtained and the merger history can be constructed
by performing Monte-Carlo realization according to the conditional
mass function. In Fig. \ref{fig:MF} mass functions by EPS and ST are shown.
One can see that the ST mass function has a larger number of massive halos.
Below we use EPS unless otherwise noted.

The merger history obtained here provides only evolution of the halo
mass and does not give spatial information such as the relative velocity
and the angular momentum of halos, which one may consider as important
ingredients to estimate the gravitaional wave emission.  Fortunately,
as we showed in \cite{2010PhRvD..82l4007I}, the initial relative velocity does
not affect the gravitational wave emission so much. Specifically,
the difference in the emitted energies is about $20\%$ between models with
zero initial relative velocity and $220~{\rm km/s}$ which is the maximum
initial relative velocity of two gravitationally-bound halos.
On the other hand, the relative angular momentum has a relatively larger
effect on the gravitational wave emission. A head-on collision emits
three times more energy than a collision where two halos initially have
a circular orbit. In the current paper, we assume head-on merger for
all the collisions so that our calculation would be over-estimation
by a factor less than three.

In constructing the merger history we take the mass resolution
$M_l = 10^{10}h^{-1}M_\odot$, that is, we ignore halos with masses less
than $M_l$ and only consider merger of two halos. The adopted time step
of the merger histories is a redshift interval of $\Delta z=0.06(1+z)$,
corresponding to the dynamical timescale of halos that collapse at redshift z.

As the second step, we sum up the gravitational waves from dark halo mergers
following the merger history obtained above, using the result of 
\cite{2010PhRvD..82l4007I} which provides the gravitational-wave spectrum
from a single merger as a function of the halo masses. This spectral
function must cover a wide range of masses and mass ratios and in
\cite{2010PhRvD..82l4007I} we found a scaling relation of the spectrum
with respect to them. The spectrum of gravitational waves from a merger
of equal-mass halos can be written as $E_{GW,0}(f,M_0)$ where $f$ and
$M_0$ are the frequency and the fiducial mass. The energy spectrum
for a given mass $M$ can be decribed by,
\begin{equation}
E_{GW}(f,M)= \left(\frac{M}{M_0}\right)^{1+\frac{7}{\alpha}}
E_{GW,0}\left(\left(\frac{M}{M_0}\right)^{1-\frac{3}{\alpha}}f,M_0\right),
\label{eq:e}
\end{equation}
where $\alpha=3.4$. The energy spectrum in the case of unequal masses,
$M_1$ and $M_2$, can be obtained by replacing $M$ in the Eq. (\ref{eq:e})
with $\sqrt{M_1M_2}$. Here we set $M_0 = 3.8 \times 10^{12}h^{-1} M_\odot$
and then the total energy is $E_{GW,0} = 5 \times 10^{46}~ h^{-1} {\rm erg}$.

\begin{figure}[t]
\includegraphics[width=80mm]{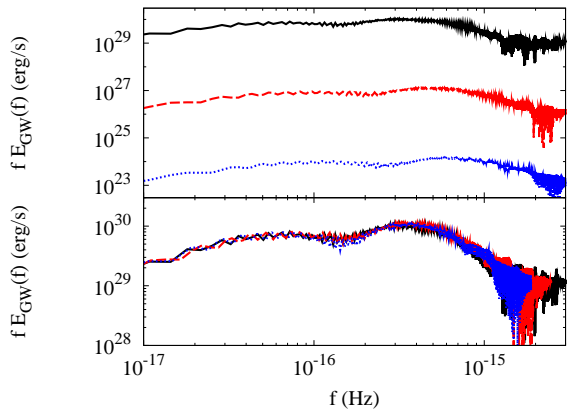}
\caption{The energy spectra of the gravitational waves 
in the logarithmic spacing for mergers of
equal-mass halos, $M/h^{-1}M_\odot = 3.8\times 10^{12}$ (solid line),
$3.8\times 10^{11}$ (dashed line) and $3.8\times 10^{10}$ (dotted).
Y-axis is the energy spectra of the gravitational waves from a merger 
in the unit of $h^{-1}{\rm erg/s}$. 
Top: the original spectra. 
Bottom: the spectra scaled by Eq. (\ref{eq:e}).}
\label{fig:fig-3-2}
\end{figure}

Here we demonstrate how valid the scaling relation Eq. (\ref{eq:e}) is.
Fig. \ref{fig:fig-3-2} shows the energy spectra of the emitted gravitational
waves in the logarithmic spacing, i.e, $fE(f)$
for three equal-mass mergers with $M = 3.8\times 10^{12}$,
$3.8\times 10^{11}$ and $3.8\times 10^{10}~h^{-1}M_\odot$. 
The top panel is the original spectra and the bottom panel is the spectra
scaled by Eq. (\ref{eq:e}) to fit with the case with
$M = 3.8\times10^{12}h^{-1}M_\odot$. In these cases, the error of
the scaling relation is less than $40\%$. 
It should be noticed that the energy spectrum has a peak 
between $10^{-17}$ to $10^{-15} {\rm Hz}$ which corresponds to 
the dynamical time scale of halo merger, $\sim 1 {\rm Gyr}$.

\begin{figure}[t]
\includegraphics[width=80mm]{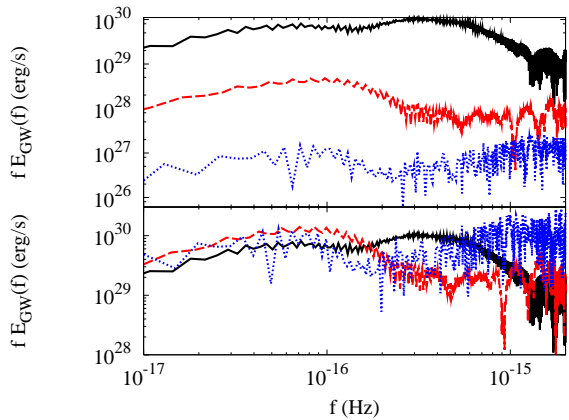}
\caption{The energy spectra of the gravitational waves 
in the logarithmic spacing for mergers of
unequal-mass halos. The solid line shows the case of a merger of equal-mass
halos $M/h^{-1}M_\odot=3.8\times 10^{12}$ shown for comparison. 
The dashed and doted lines show the cases of unequal masses
(dashed line:$M/h^{-1}M_\odot=3.8\times 10^{11}$ and $3.8\times 10^{12}$, 
dotted line: $M/h^{-1}M_\odot=3.8\times 10^{10}$ and $3.8\times 10^{12}$).
Y-axis is the energy spectra of the gravitational waves from a merger 
in the unit of $h^{-1}{\rm erg/s}$. 
Top: the original spectra. 
Bottom: the spectra scaled by Eq. (\ref{eq:e}).}
\label{fig:fig-2-2}
\end{figure}

Fig. \ref{fig:fig-2-2} is a demonstration for mergers of unequal-mass halos.
Here we fix the higher mass to $3.8 \times 10^{12} h^{-1} M_\odot$
and vary the mass ratio as $1:1$, $1:1/10$ and $1:1/100$.
The top panel in Fig. \ref{fig:fig-2-2} shows the original spectra and
the bottom panel shows the spectra scaled by Eq. (\ref{eq:e}).
Although the error of the scaling relation is relatively large at high
frequencies ($> 10^{-16}~{\rm Hz}$), it is still reasonable at low
frequencies where most of the energy is emitted.
Thus, we use Eq. (\ref{eq:e}) as a reasonable scaling relation
for both equal- and unequal-mass mergers.

The energy density of the GWB at redshift $z_i$ can be calculated as,
\begin{equation}
\rho_{GW}(f, z_i)=\frac{1}{V_{\rm com}}\sum_{N_i}E^{(N_i)}_{GW}(f),  
\label{eq:e1}
\end{equation}
where $V_{\rm com}$ is a comoving volume and $N_i$ represents
the $N_i$-th merger in the $i$-th redshift bin. It should be noted that,
due to the expansion of the universe, the frequency and
the energy density of the gravitational waves are redshifted as
$f\propto 1/(1+z)$ and $\rho_{GW}\propto 1/(1+z)^4$, respectively.
Therefore, the energy density $\rho_{GW}$ at $z=0$ can be written as,
\begin{equation}
\rho_{GW}(\tilde{f}, z=0)
= \frac{1}{V_{\rm phys}}
  \sum_{i} \sum_{N_i}\frac{1}{1+z_i}E^{(N_i)}_{GW}(\tilde{f}),
\label{eq:density}
\end{equation}
where $\tilde{f}$ and $V_{\rm phys}$ show the redshifted frequency
and a physical volume, respectively. Finally, the density parameter of
the GWB, $\Omega_{GW}$, is defined as,
\begin{equation}
\Omega_{GW}(\tilde{f})
\equiv
\frac{1}{\rho_{c,0}c^2}
\left| \frac{d\rho_{GW}(\tilde{f}, z=0)}{d\ln\tilde{f}} \right| ,
\label{eq:omega}
\end{equation}
where $\rho_{c,0}$ and $c$ show the critical density of the universe
at present and the speed of light, respectively.

\section{Results}
\label{results}

\begin{figure}[t]
\includegraphics[width=85mm]{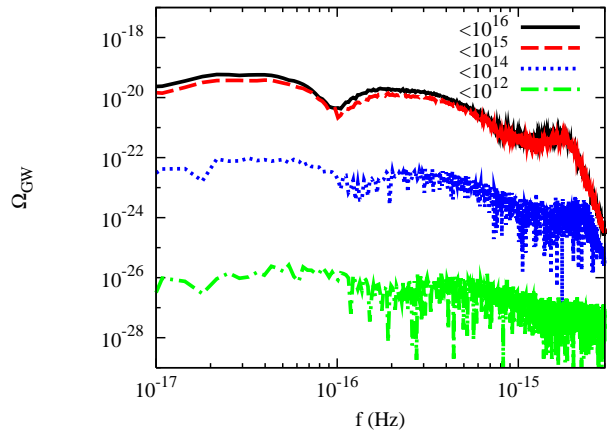}
\caption{The GWB spectra for each mass range.
The solid, dashed and dotted lines are mass range of 
$M/h^{-1}M_\odot \leq 10^{16}$, $10^{15}$, $10^{14}$ 
and $10^{12}$, respectively. }
\label{fig:fig-1}
\end{figure}

We obtain the spectrum of the GWB by using the method described 
in the previous section. First we show the dependence of the result
with respect to the upper-cutoff mass $M_{\rm cut}$ of halos.
Fig. \ref{fig:fig-1} represents the GWB spectra for several mass cutoffs,
$10^{16} h^{-1} M_{\odot}$, $10^{15} h^{-1} M_{\odot}$,
$10^{14} h^{-1} M_{\odot}$ and $10^{12} h^{-1} M_{\odot}$.
The emitted energy reaches $\Omega_{GW} \sim 5 \times 10^{-20}$
for $M_{\rm cut} \leq 10^{16} h^{-1} M_{\odot}$ and the spectral
shape is very similar to that of a single merger shown in
Fig. \ref{fig:fig-3-2}. The difference in the spectra is very small
for the cases with $M_{\rm cut} = 10^{15} h^{-1} M_{\odot}$ and
$10^{16} h^{-1} M_{\odot}$. This can  be easily understood by the shape
of the mass function in Fig. \ref{fig:MF}, where the number of halos
steeply decreases above $10^{15} h^{-1} M_\odot$. On the other hand,
the difference is about two orders of magnitude between the cases
with $M_{\rm cut} = 10^{14} M_{\odot}$ and $10^{15} h^{-1} M_{\odot}$,
which implies that most of GWB energy is contributed from mergers
of massive halos. Because the GWB spectrum is saturated at
$M_{\rm cut} = 10^{15} h^{-1} M_{\odot}$, we take
$10^{15} h^{-1} M_{\odot}$ as a cutoff mass hereafter.

\begin{figure}[t]
\includegraphics[width=85mm]{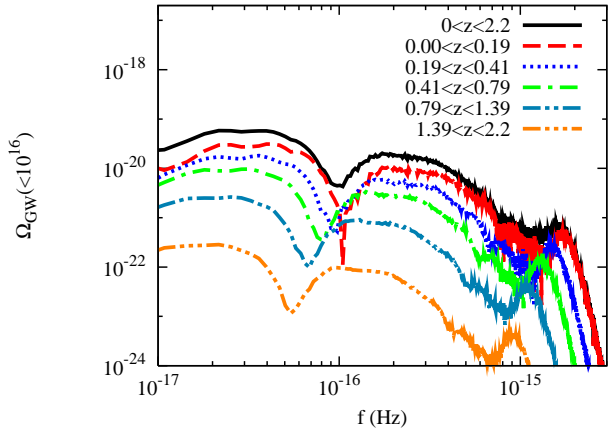}
\caption{Contributions of several redshift ranges to the GWB spectrum.
}
\label{fig:fig-redshift}
\end{figure}

\begin{figure}[t]
\includegraphics[width=85mm]{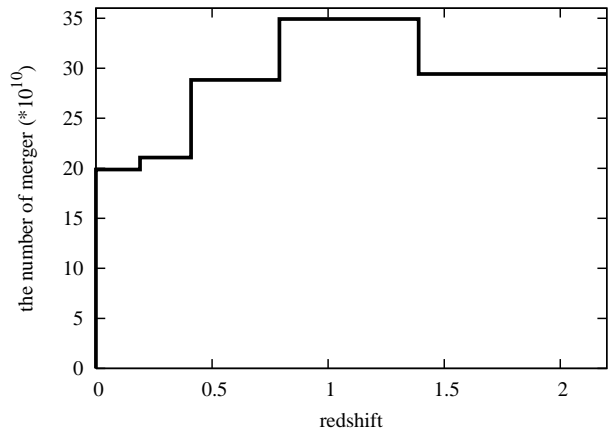}
\caption{The total number of merger at each redshift. 
The assumed volume is $(6~h^{-1}{\rm Gyr})^3$
correspoding to hubble scale.}
\label{fig:fig-redshift2}
\end{figure}

\begin{figure}[t]
\includegraphics[width=85mm]{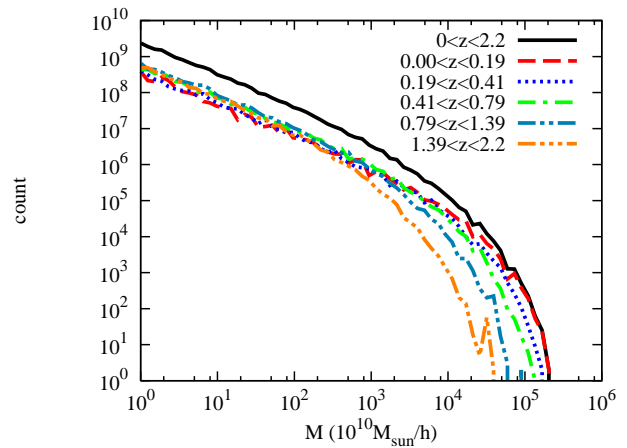}
\caption{The number of mergers at each redshift with respect to mass
that is higher mass of two merger haloes. The assumed volume is
$(6~h^{-1}{\rm Gyr})^3$ correspoding to hubble scale.
}
\label{fig:fig-N}
\end{figure}

Next we investigate the relative contributions of redshift ranges to
the GWB spectrum. In Fig. \ref{fig:fig-redshift}, the contributions
from several redshift bins are shown. The widths of the bins are determined
so that each bin corresponds to the same cosmic time interval of
$2~{\rm Gyr/h}$. Each spectrum in Fig. \ref{fig:fig-redshift} represents
the GWB emitted by mergers in each redshift range. Due to the redshift
of gravitational wave, contributions from larger $z$ bins have peaks
at lower frequencies. We find $46\%$ of the total energy of
the gravitational waves comes from the redshift interval
$0\leq z \leq 0.19$, while the gravitational waves from the
$0.19 \leq z \leq 0.41$ and $0.41 \leq z \leq 0.79$ contribute to
the GWB spectrum by about $26\%$ and $15\%$, respectively.
Thus, about $90\%$ of the GWB comes from the redshift interval
$0 \leq z \leq 0.79$.

To understand this behavior, we show the merger rate below.
Fig. \ref{fig:fig-redshift2} represents the total number of halo
mergers for each redshift bin of Fig. \ref{fig:fig-redshift}.
The total number has a peak at $0.79 < z < 1.39$ and decreases
toward lower $z$. However, as can be seen in Fig. \ref{fig:fig-N}
which represents the merger rate as a function of mass, mergers
of massive halos takes place more frequent for lower $z$. This would be
natural in the hierarchical scenario of structure formation.
Because the energy of emitted graviational waves increases
substantially with the halo masses (see Eq. (\ref{eq:e})), the dominant
contribution to the resultant spectrum of GWB comes from
the lowest-redshift bin although the total number of mergers
is relatively small. Here we remark that we have checked that
the contribution from $z > 2.2$ is negligible as expected.

\begin{figure}[t]
\includegraphics[width=80mm]{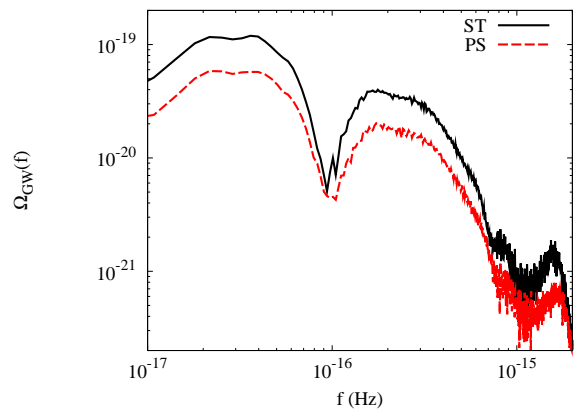}
\caption{The spectrum by using ST(solid) and EPS(dashed). }
\label{fig:ST-PS}
\end{figure}

Finally, we compare the spectra of the GWB calculated using EPS and ST
mass functions in Fig. \ref{fig:fig-1}. Here we took the cutoff mass
$M_{\rm cut} = 10^{16} h^{-1} M_{\odot}$. The total energy of the GWB
for ST case is larger than that for EPS case by a factor of two and
reaches $\Omega_{GW} \sim 10^{-19}$. This is because massive halos
are more abundant for ST than EPS.

\section{Summary and Discussion}
\label{summary}

In this paper, we calculated the spectrum of gravitational wave
background originating from dark halo mergers by a quasi-analytic method. 
First, we constructed merger histories by Monte-Carlo realizations for 
a mass function (Extended Press-Schechter formalism or the Sheth \&
Tormen formalism). Then we summed up the energy spectra from halo
mergers following the merger history, using the result of 
\cite{2010PhRvD..82l4007I} which provides the gravitational-wave
spectrum from a single merger as a function of the halo masses.
We found that the energy density reaches
$\Omega_{GW} \sim 5 \times 10^{-20}$ and that the dominant contribution
comes from mergers of massive halos $10^{14}-10^{15} h^{-1} M_{\odot}$
at relatively low redshifts $z < 0.19$. We gave an interpretation of
the relative importance of ranges of halo masses and redshifts showing
the merger rates as a function of mass and redshift. We also compared
the GWB spectra obtained by EPS to ST mass functions and found that
the latter case has larger energy by a factor of two.

Finally, we discuss the observability of the GWB from halo mergers.
Stochastic gravitational wave background converts 
the E-mode polarization of CMB into the B-mode through gravitational 
lensing between the observer and the last scattering. 
Thus observation of B-mode would be useful to probe GWs after last scattering.
The energy of the GWB is at most $\Omega_{GW}\sim 10^{-19}$ according
to our calculations. This corresponds to inflationary gravitational waves
with the tensor-to-scalar ratio $r\sim 10^{-4}$ \cite{2009PhRvD..79j3501K}.
It may be possible to detect them through B-mode polarization of CMB
if the tensor-to-scalar ratio is lower than above the value.
In this case, our result would be useful to probe the process of
structure formation.

\begin{acknowledgements}

TI is supported by JSPS. 
This work is supported in part by JSPS
Grant-in-Aid for the Global COE programs, ``Quest for
Fundamental Principles in the Universe: from Particles to
the Solar System and the Cosmos'' at Nagoya University.
KT is supported by Grand-in-Aid for Scientific Research No.~23740179.
NS is supported by Grand-in-Aid for Scientific Research No.~22340056
and 18072004.
The authors acknowledge Kobayashi-Maskawa Institute for the Origin of 
Particles and the Universe, Nagoya University for providing computing 
resources useful in conducting the research reported in this paper. 
This research has also been supported in part by World Premier
International Research Center Initiative,
MEXT, Japan.

\end{acknowledgements}

\bibliography{GW}

\begin{thebibliography}{16}
\expandafter\ifx\csname natexlab\endcsname\relax\def\natexlab#1{#1}\fi
\expandafter\ifx\csname bibnamefont\endcsname\relax
  \def\bibnamefont#1{#1}\fi
\expandafter\ifx\csname bibfnamefont\endcsname\relax
  \def\bibfnamefont#1{#1}\fi
\expandafter\ifx\csname citenamefont\endcsname\relax
  \def\citenamefont#1{#1}\fi
\expandafter\ifx\csname url\endcsname\relax
  \def\url#1{\texttt{#1}}\fi
\expandafter\ifx\csname urlprefix\endcsname\relax\def\urlprefix{URL }\fi
\providecommand{\bibinfo}[2]{#2}
\providecommand{\eprint}[2][]{\url{#2}}

\bibitem[{\citenamefont{{Sato}}(1981)}]{1981MNRAS.195..467S}
\bibinfo{author}{\bibfnamefont{K.}~\bibnamefont{{Sato}}},
  \bibinfo{journal}{MNRAS} \textbf{\bibinfo{volume}{195}}, \bibinfo{pages}{467}
  (\bibinfo{year}{1981}).

\bibitem[{\citenamefont{Guth}(1981)}]{PhysRevD.23.347}
\bibinfo{author}{\bibfnamefont{A.~H.} \bibnamefont{Guth}},
  \bibinfo{journal}{Phys. Rev. D} \textbf{\bibinfo{volume}{23}},
  \bibinfo{pages}{347} (\bibinfo{year}{1981}).

\bibitem[{\citenamefont{Albrecht and Steinhardt}(1982)}]{PhysRevLett.48.1220}
\bibinfo{author}{\bibfnamefont{A.}~\bibnamefont{Albrecht}} \bibnamefont{and}
  \bibinfo{author}{\bibfnamefont{P.~J.} \bibnamefont{Steinhardt}},
  \bibinfo{journal}{Phys. Rev. Lett.} \textbf{\bibinfo{volume}{48}},
  \bibinfo{pages}{1220} (\bibinfo{year}{1982}).

\bibitem[{\citenamefont{{Linde}}(1982)}]{1982PhLB..108..389L}
\bibinfo{author}{\bibfnamefont{A.~D.} \bibnamefont{{Linde}}},
  \bibinfo{journal}{Physics Letters B} \textbf{\bibinfo{volume}{108}},
  \bibinfo{pages}{389} (\bibinfo{year}{1982}).

\bibitem[{\citenamefont{{Kuroyanagi} et~al.}(2009)\citenamefont{{Kuroyanagi},
  {Chiba}, and {Sugiyama}}}]{2009PhRvD..79j3501K}
\bibinfo{author}{\bibfnamefont{S.}~\bibnamefont{{Kuroyanagi}}},
  \bibinfo{author}{\bibfnamefont{T.}~\bibnamefont{{Chiba}}}, \bibnamefont{and}
  \bibinfo{author}{\bibfnamefont{N.}~\bibnamefont{{Sugiyama}}},
  \bibinfo{journal}{\prd} \textbf{\bibinfo{volume}{79}},
  \bibinfo{pages}{103501} (\bibinfo{year}{2009}), \eprint{0804.3249}.

\bibitem[{\citenamefont{{Mollerach} et~al.}(2004)\citenamefont{{Mollerach},
  {Harari}, and {Matarrese}}}]{2004PhRvD..69f3002M}
\bibinfo{author}{\bibfnamefont{S.}~\bibnamefont{{Mollerach}}},
  \bibinfo{author}{\bibfnamefont{D.}~\bibnamefont{{Harari}}}, \bibnamefont{and}
  \bibinfo{author}{\bibfnamefont{S.}~\bibnamefont{{Matarrese}}},
  \bibinfo{journal}{\prd} \textbf{\bibinfo{volume}{69}},
  \bibinfo{pages}{063002} (\bibinfo{year}{2004}),
  \eprint{arXiv:astro-ph/0310711}.

\bibitem[{\citenamefont{{Ananda} et~al.}(2007)\citenamefont{{Ananda},
  {Clarkson}, and {Wands}}}]{2007PhRvD..75l3518A}
\bibinfo{author}{\bibfnamefont{K.~N.} \bibnamefont{{Ananda}}},
  \bibinfo{author}{\bibfnamefont{C.}~\bibnamefont{{Clarkson}}},
  \bibnamefont{and} \bibinfo{author}{\bibfnamefont{D.}~\bibnamefont{{Wands}}},
  \bibinfo{journal}{\prd} \textbf{\bibinfo{volume}{75}},
  \bibinfo{pages}{123518} (\bibinfo{year}{2007}), \eprint{arXiv:gr-qc/0612013}.

\bibitem[{\citenamefont{{Baumann} et~al.}(2007)\citenamefont{{Baumann},
  {Steinhardt}, {Takahashi}, and {Ichiki}}}]{2007PhRvD..76h4019B}
\bibinfo{author}{\bibfnamefont{D.}~\bibnamefont{{Baumann}}},
  \bibinfo{author}{\bibfnamefont{P.}~\bibnamefont{{Steinhardt}}},
  \bibinfo{author}{\bibfnamefont{K.}~\bibnamefont{{Takahashi}}},
  \bibnamefont{and} \bibinfo{author}{\bibfnamefont{K.}~\bibnamefont{{Ichiki}}},
  \bibinfo{journal}{\prd} \textbf{\bibinfo{volume}{76}},
  \bibinfo{pages}{084019} (\bibinfo{year}{2007}),
  \eprint{arXiv:hep-th/0703290}.

\bibitem[{\citenamefont{{Pritchard} and
  {Kamionkowski}}(2005)}]{2005AnPhy.318....2P}
\bibinfo{author}{\bibfnamefont{J.~R.} \bibnamefont{{Pritchard}}}
  \bibnamefont{and}
  \bibinfo{author}{\bibfnamefont{M.}~\bibnamefont{{Kamionkowski}}},
  \bibinfo{journal}{Annals of Physics} \textbf{\bibinfo{volume}{318}},
  \bibinfo{pages}{2} (\bibinfo{year}{2005}), \eprint{arXiv:astro-ph/0412581}.

\bibitem[{\citenamefont{{Lacey} and {Cole}}(1993)}]{1993MNRAS.262..627L}
\bibinfo{author}{\bibfnamefont{C.}~\bibnamefont{{Lacey}}} \bibnamefont{and}
  \bibinfo{author}{\bibfnamefont{S.}~\bibnamefont{{Cole}}},
  \bibinfo{journal}{MNRAS} \textbf{\bibinfo{volume}{262}}, \bibinfo{pages}{627}
  (\bibinfo{year}{1993}).

\bibitem[{\citenamefont{{Lacey} and {Cole}}(1994)}]{1994MNRAS.271..676L}
\bibinfo{author}{\bibfnamefont{C.}~\bibnamefont{{Lacey}}} \bibnamefont{and}
  \bibinfo{author}{\bibfnamefont{S.}~\bibnamefont{{Cole}}},
  \bibinfo{journal}{MNRAS} \textbf{\bibinfo{volume}{271}}, \bibinfo{pages}{676}
  (\bibinfo{year}{1994}), \eprint{arXiv:astro-ph/9402069}.

\bibitem[{\citenamefont{{Inagaki} et~al.}(2010)\citenamefont{{Inagaki},
  {Takahashi}, {Masaki}, and {Sugiyama}}}]{2010PhRvD..82l4007I}
\bibinfo{author}{\bibfnamefont{T.}~\bibnamefont{{Inagaki}}},
  \bibinfo{author}{\bibfnamefont{K.}~\bibnamefont{{Takahashi}}},
  \bibinfo{author}{\bibfnamefont{S.}~\bibnamefont{{Masaki}}}, \bibnamefont{and}
  \bibinfo{author}{\bibfnamefont{N.}~\bibnamefont{{Sugiyama}}},
  \bibinfo{journal}{\prd} \textbf{\bibinfo{volume}{82}},
  \bibinfo{pages}{124007} (\bibinfo{year}{2010}), \eprint{1011.5554}.

\bibitem[{\citenamefont{{Komatsu} et~al.}(2011)\citenamefont{{Komatsu},
  {Smith}, {Dunkley}, {Bennett}, {Gold}, {Hinshaw}, {Jarosik}, {Larson},
  {Nolta}, {Page} et~al.}}]{2011ApJS..192...18K}
\bibinfo{author}{\bibfnamefont{E.}~\bibnamefont{{Komatsu}}},
  \bibinfo{author}{\bibfnamefont{K.~M.} \bibnamefont{{Smith}}},
  \bibinfo{author}{\bibfnamefont{J.}~\bibnamefont{{Dunkley}}},
  \bibinfo{author}{\bibfnamefont{C.~L.} \bibnamefont{{Bennett}}},
  \bibinfo{author}{\bibfnamefont{B.}~\bibnamefont{{Gold}}},
  \bibinfo{author}{\bibfnamefont{G.}~\bibnamefont{{Hinshaw}}},
  \bibinfo{author}{\bibfnamefont{N.}~\bibnamefont{{Jarosik}}},
  \bibinfo{author}{\bibfnamefont{D.}~\bibnamefont{{Larson}}},
  \bibinfo{author}{\bibfnamefont{M.~R.} \bibnamefont{{Nolta}}},
  \bibinfo{author}{\bibfnamefont{L.}~\bibnamefont{{Page}}},
  \bibnamefont{et~al.}, \bibinfo{journal}{ApJS} \textbf{\bibinfo{volume}{192}},
  \bibinfo{eid}{18} (\bibinfo{year}{2011}), \eprint{1001.4538}.

\bibitem[{\citenamefont{{Somerville} and {Kolatt}}(1999)}]{1999MNRAS.305....1S}
\bibinfo{author}{\bibfnamefont{R.~S.} \bibnamefont{{Somerville}}}
  \bibnamefont{and} \bibinfo{author}{\bibfnamefont{T.~S.}
  \bibnamefont{{Kolatt}}}, \bibinfo{journal}{MNRAS}
  \textbf{\bibinfo{volume}{305}}, \bibinfo{pages}{1} (\bibinfo{year}{1999}),
  \eprint{arXiv:astro-ph/9711080}.

\bibitem[{\citenamefont{{Sheth} and {Tormen}}(1999)}]{1999MNRAS.308..119S}
\bibinfo{author}{\bibfnamefont{R.~K.} \bibnamefont{{Sheth}}} \bibnamefont{and}
  \bibinfo{author}{\bibfnamefont{G.}~\bibnamefont{{Tormen}}},
  \bibinfo{journal}{MNRAS} \textbf{\bibinfo{volume}{308}}, \bibinfo{pages}{119}
  (\bibinfo{year}{1999}), \eprint{arXiv:astro-ph/9901122}.

\bibitem[{\citenamefont{{Jenkins} et~al.}(2001)\citenamefont{{Jenkins},
  {Frenk}, {White}, {Colberg}, {Cole}, {Evrard}, {Couchman}, and
  {Yoshida}}}]{2001MNRAS.321..372J}
\bibinfo{author}{\bibfnamefont{A.}~\bibnamefont{{Jenkins}}},
  \bibinfo{author}{\bibfnamefont{C.~S.} \bibnamefont{{Frenk}}},
  \bibinfo{author}{\bibfnamefont{S.~D.~M.} \bibnamefont{{White}}},
  \bibinfo{author}{\bibfnamefont{J.~M.} \bibnamefont{{Colberg}}},
  \bibinfo{author}{\bibfnamefont{S.}~\bibnamefont{{Cole}}},
  \bibinfo{author}{\bibfnamefont{A.~E.} \bibnamefont{{Evrard}}},
  \bibinfo{author}{\bibfnamefont{H.~M.~P.} \bibnamefont{{Couchman}}},
  \bibnamefont{and}
  \bibinfo{author}{\bibfnamefont{N.}~\bibnamefont{{Yoshida}}},
  \bibinfo{journal}{MNRAS} \textbf{\bibinfo{volume}{321}}, \bibinfo{pages}{372}
  (\bibinfo{year}{2001}), \eprint{arXiv:astro-ph/0005260}.

\end{thebibliography}

\end{document}